# THE OBSERVATION OF FORMATION AND ANNIHILATION OF SOLITONS AND STANDING STRAIN WAVE SUPERSTRUCTURES IN A TWO-DIMENSIONAL COLLOIDAL CRYSTAL

Yu-Hang Chui[1*], Surajit Sengupta[2], Ian K. Snook[3] and Kurt Binder[1]

[1]Institute of Physics, Johannes-Gutenberg University
D-55099 Mainz, Staudinger Weg 7, Germany

[2] Centre for Advanced Materials, Indian Association for the Cultivation of Science, Jadavpur, Kolkata 700032, India and
S. N. Bose National Centre for Basic Science, Block JD, Sector III,
Salt Lake, Kolkata 70098, India

[3]Applied Physics, School of Applied Science, RMIT University, B. O. Box 2476V, 3001 Victoria, Australia

Corresponding author, email address: yuhang@graduate.hku.hk

## Abstract

Confining a colloidal crystal within a long narrow channel produced by two parallel walls can be used to impose a meso-scale superstructure of a predominantly mechanical elastic character [Chui et al., EPL **2008**, *83*, 58004]. When the crystal is compressed in the direction perpendicular to the walls, we obtain a structural transition when the number of rows of particles parallel to the walls decreases by one. All the particles of this vanishing row are distributed throughout the crystal. If the confining walls are structured (say with a corrugation along the length of the walls), then these extra particles are distributed neither uniformly nor randomly; rather, defect structures are created along the boundaries resembling "soliton staircases", inducing a non-uniform strain pattern within the crystal. Here we study the conditions of stability, formation and annihilation of these solitons using a coarse grained description of the dynamics. The processes are shown by comparing superimposed configurations as well as molecular animations obtained from our simulations. Also the corresponding normal and shear stresses during the transformation are calculated. A study of these dynamical processes should be useful for controlling strain wave superstructures in the self-assembly of various nano- and meso scaled particles.

## Introduction

As is well known, “soliton staircases” are periodic patterns of localized defects that arise from the lack of commensurability of a (one-dimensional) crystal with a periodic potential that it is exposed to. In our previous study, we have shown that confinement can be used to impose a controllable mesoscopic superstructure of a predominantly mechanical elastic character on a crystal [1]. Due to an interplay of the particle density of the system and the width *D* of a confining channel, ``soliton staircases''[2] can be created along both parallel confining boundaries, that give rise to standing strain waves in the entire crystal. This new type of mesophase is reminiscent of charge density waves[3] and spin density waves[4] in crystals, but was demonstrated for a model of a two-dimensional colloidal crystal, occurring hence on much larger length scales. Related phenomena could also occur for dusty plasmas[5], lattices of spherical block copolymer micelles under confinement[6], superstructures of small molecules or atoms adsorbed on stepped surfaces[7], and superlattices of nano- and meso-scaled particles[8]. A related phenomenon of standing strain waves induced by a boundary condition might occur in epitaxially grown thin films with lattice misfit such as Fe films on W (110), where a periodic structure of misfit dislocations at the Fe-W interface may cause a modulation that decreases with the distance from the interface in the thin film[9]. The present work also deals with a Monte Carlo (MC) model for a confined colloidal crystal in two dimensions, for which we investigate the stability, formation and annihilation of solitons caused by varying the misfit in the distance between two corrugated walls. We use a local MC algorithm to model the dynamics of suspended colloidal particles, executing random Brownian motion due to collisions with the small solvent particles[10-12]. We have neglected hydrodynamic interactions as we are not concerned with the small-scale details (happening on very short time scales) of the mechanism but only the coarse- grained dynamics at large scales where, in the absence of external fields, it has been shown that hydrodynamic effects do not alter the qualitative form of the long time dynamics[13-15].

With increasing misfit (i.e. strain) we observed that the stress increases up to some critical value, where a transition occurs that *reduces the number of crystalline rows parallel to the boundaries by one*. At constant particle number, the extra particles of the row that disappears are distributed in the system such that a ``soliton staircase'' [2] is created along the walls, accompanied by a pattern of standing strain waves in the crystal. Reduction of the misfit in the strained colloidal crystal will cause the inverse transition which *increases the number of crystalline rows by one*. We believe the understanding of non-equilibrium dynamics will help to give a better control of the standing strain wave pattern in two-dimensional crystals.

## Model and simulation protocol

The colloidal particles are described as point particles interacting with a potential $V(r) = \varepsilon(\sigma / r)^{16}$, where $\varepsilon$ sets the energy scale, $\sigma$ sets the distance scale, and $r$ denotes the interparticle distance. At low enough temperatures $T$ this system forms a crystal with a triangular lattice structure, where the lattice parameter $a$ is related to the chosen density $\rho$ via $a^2 = 2/(\sqrt{3}\rho)$. Certainly, for such systems with inverse power law potentials, $T$ and $\rho$ are not independent control parameters, in our case it is only the combination $\rho(\varepsilon / k_B T)^{1/6}$ that matters[17,18]. Thus, choosing length units such that $\rho = 1.05$ one finds that the (presumably continuous) melting transition of the crystal occurs at $k_B T_m / \varepsilon \approx 1.35$ [19].

Following Ricci et al.[17], we create a confinement potential commensurate with this lattice structure by putting two rows of frozen particles at either side of the crystal, which in our case has $n_x$ rows containing $n_y$ particles each, so that the crystal in the case where there is no misfit has the linear dimensions $L_y = n_y a$ and $D = n_x a\sqrt{3}/2$. These rows of periodically arranged frozen particles create a periodic potential (with periodicity given by the lattice spacing a) acting on the mobile particles in the strip, thus stabilizing a crystalline structure with this periodicity. However, by choosing a smaller distance between the frozen rows on both sides of this crystalline strip we can enforce a misfit, such that $D = (n_x-\Delta)a\sqrt{3}/2$. This equation defines the misfit parameter *D*.

Here we have used a Monte Carlo (MC) method to give a coarse-grained description of the dynamics of stability, formation and annihilation of solitons. The method used generated particle positions by the Metropolis Monte Carlo process which moved each particle at each MC time step with a maximum displacement of 0.1 lattice spacing, *a*. Particles are selected at random for such trial moves, and acceptance or rejection of the move is controlled by the standard Metropolis transition probability.

In effect, this method is similar to a Brownian Dynamics (BD) method but ignoring hydrodynamic interactions to investigate the broad, coarse-grained picture of the dynamics of the defect structures, the "solitons" that will be characterized below. Thus, we are concerned with the large-scale features of the stability, formation and annihilation of the solitons.

Note that while hydrodynamics interactions alters the qualitative dynamics of the colloidal particles at times that are short relative to the time for a particle to equilibrate its position in the "cage" formed by its neighboring particles, at long time scales, in the absence of an imposed external field[15], the effect of the hydrodynamic forces can fully be accounted for by a renormalization of the effective time scale[13, 14]. Hence, mode coupling theory (MCT)[13,14] without hydrodynamic forces provides a quantitatively accurate and highly nontrivial description of all dynamic correlation functions of the system, after the "microscopic time scale" is fixed[13,14].

Further, the MC method has been used to describe the slow dynamics of dense polymer melts where it successfully predicts the Rouse to Reptation crossover as accurately as molecular dynamics and in agreement with experiments[20].

In our previous work we determined the variation of stress with strain (misfit)[1]. Higher and higher stress is built up while compressing the system by increasing the misfit (from $\Delta$ = 0 to $\Delta$ = 2.0) in the colloidal crystals. This is done in steps of $\Delta = 0.25$ (see Figure 1). We first choose the positions of the wall atoms such that $\Delta = 0$, with an initial condition of a perfect triangular lattice structure, and equilibrate the system at $k_b T / \varepsilon = 1$ by standard Monte Carlo methods[21]. Periodic boundary conditions are used in y direction, and typically linear dimensions are chosen such that we have $n_x$=30 rows with 108 particles per row. Then we move the rows with the wall atoms closer to each other, in order to create a misfit with $\Delta = 0.25$. This is done by generating a starting configuration compatible with this reduced distance between two walls, by uniform rescaling of all particle distances perpendicular to the walls using the appropriate factor. After moving the wall atoms closer to each other, we equilibrate the system (typical equilibration time were the order of 8 million Monte Carlo steps per particle), then increase the misfit to $\Delta = 0.5$ and re-run the simulation, and so forth. Recording the stress $\sigma = \sigma_{xx} - \sigma_{yy}$ [22] from the computation of the virial tensor[12], an almost linear increase of the stress up to a maximum value at about $\Delta = 2$ is found (filled square in the lower part of figure 1), where an abrupt first-order-like transition to a slightly negative value of the stress occurs. Here a crystal row disappears; a detailed analysis of the time-dependence of the process (see also the animation in the supplementary material) suggests that the mechanism for this transition happens via the nucleation and subsequent annihilation of a pair of dislocations with opposite Burger's vectors perpendicular to the walls.

The kinetics of this transition is the phenomenon which we wish to describe in the present paper. Increasing the misfit further up to $\Delta = 3$, the stress increases again (Fig. 1).

Since we have found[1] (see also next section) that the transition at $\Delta = 2$ where one row with $n_y$ particles disappeared corresponds to the formation of an almost periodic defect structure, we have also created a strictly periodic structure with $n_x - 1$ rows (i.e. 29 rows in this study). Since it was found[1] that the rows adjacent to the walls (open circles in the upper part of Fig. 1) do not host any extra particles from that disappeared row, we use a structure with $n_y + n_y / (n_x - 3)$ particles per row as an initial condition (open circles in the lower part of Fig. 1).

To study the reverse process (soliton annihilation) we start out at $\Delta = 3$, the system is equilibrated with this initial condition, and then $\Delta$ is reduced in steps of 0.25 and re-equilibrated at $\Delta = 1.5$. As a result, a transition back from $n_x$-1 to $n_x$ rows is obtained (open triangles in the lower part of Fig.1). During these structural transitions that occurs after varying the misfit, we observed and recorded the kinetics and the variation of normal stress and shear stress in the colloidal crystal.

As a test, we also simulate the system with *D* smaller than 1.5. Then the stresses before and after soliton annihilation were also plotted. With *D* larger than 1.5, the soliton annihilation does not occur.

## Results on the transition kinetics

As described above, compression of the two-dimensional crystal strip leads to a transition $n_x \rightarrow n_x$-1 in the number of rows, and the $n_y$ particles of this disappearing row are distrbuted over the $n_x$ – 3 inner rows of the strip. If there were no effect due to the periodic potential of the two confining walls, one would expect that the strip forms a crystal with a lattice spacing a' in y direction that is reduced in order to accommodate more particles $a' = a / [1+1/( n_x – 3)]$. However, a lattice with this reduced lattice constant is incommensurate with the periodic boundary potential created by the fixed wall particles, which remains at the original periodicity.

This conflict of periodicities at the walls is reminiscent of the problem of a harmonic crystal in one-dimension exposed to a periodic potential: if the periodicity of this crystal coincides with the periodicity of the potential, all particles will sit in the potential walls (Fig. 2a). However, if the number of particles in the harmonic chains exceeds the number of potential wells slightly, the crystal contains defects, so-called "solitons" where a few particles are forced to be at positions different from the potential minima (Fig.2b). The minimum energy configuration is then a periodic arrangement of these defects, the so-called "soliton staircase"[2]. In between the solitons the particles stay eventually in the potential wells.

Of course, the choice of which the particles sits on top of a potential well (full dot in Fig. 2b) is arbitrary, and due to this degeneracy the defect structure shown in Fig.2b can be translated along the y axis (without energy cost when the soliton lattice moves as a whole, with little energy cost when an individual soliton moves, as long as the distance to the neighboring soliton is large). While in the mechanical problem of Fig.2a, b the motion of solitons has the character of travelling waves, in a colloidal system due to friction of colloidal particles caused by the solvent fluid only a diffusive motion of solitons is expected.

When we now consider superimposed configurations of the particles (Fig. 2c), we see that particles staying in potential wells show up as irregular black dots (the "size" of these dots gives a measure of the typical mean square displacement of the particles). Particles in solitons, however, show up as dark stripes smeared out a distance of the overall lattice spacing in y-direction, due to the lateral diffusion of these defects (Fig. 2c). However, we can also see that these defects are not localized in single row, but are extended over several adjacent rows. Far away from the walls where the periodic potential is created, however, all the particles are again localized to lattice positions (but now the lattice spacing in y-direction indeed has a smaller value $d < a$).

In the following, we shall explain how these defect structures are formed during the transition $n_x \rightarrow n_x$-1, as well as how these defect structures disappear again for the reverse process, $n_x$-1 $\rightarrow n_x$.

***(i) Non-equilibrium dynamics in two-dimensional colloidal crystal***

We observe the kinetics in the two-dimensional colloidal crystal from superimposed configurations of the particle positions. Figure 3 shows the animation corresponding to the formation of solitons with a misfit with $\Delta = 2.0$ and the sliding transitions which were observed between two structured walls. Each picture shows a superposition of 100 individual configurations, which were taken every 100 MC steps. In Figure 3a, a sliding centre (which is the two-dimensional analog of a gliding plane for dislocation motion in three dimensions) moves downward progressively from the soliton on the top of the system (indicated by solid arrow) to the bottom, and then a new soliton is formed beside the wall, which is shown in Figure 3b, indicated by an ellipse encircling the defect. Also in Figure 3b, the sliding centre moves upward to the top arrow and the other soliton is formed. Figure 3c and 3d show that more and more solitons are formed during the zigzag motion of the sliding centre in the structural transition. These phenomena are analogous to transport phenomena in three-dimensional crystals during plastic deformation[23].

From the animation in Figure 4, we observe the structural transformation back from $n_x$-1 to $n_x$ rows during the annihilation of solitons. A similar sliding transition is also found but the sliding centre moves away from the annihilated soliton to the opposite soliton beside the opposite wall of the system (Figure 4b and 4c). After the zigzag motion, solitons are deformed and then annihilated in the colloidal crystal, and finally the whole structural transformation is completed and the number of rows of colloidal particles is increased by one.

From these observations one can conclude that several competing local structures representing local minima in the free energy landscape are available to the confined solid during the layer transformation. These structures, being more or less degenerate in energy are visited by the system over long time scales. At shorter time scales, on the other hand, the system prefers to stay within any one of these local minima. Similar phenomena have been observed in hard disc particles confined within smooth walls[22, 24] and appear to be generic for such systems.

### *(ii) Stress and shear stress during structural transformation*

Other than the observable dynamics, how the system responds mechanically to the structural transformation is also interesting. Thus we calculate the normal stress and shear stress of the system according to the fluctuation formalism proposed by Farago and Kantor[25]:

$$\sigma_{ij} = \frac{1}{V}\frac{\partial F}{\partial \eta_{ij}}\bigg|_{\{\eta\}=\{0\}} = \frac{1}{V}\langle \sum_{\langle\alpha\beta\rangle} \phi'(R^{\alpha\beta})\frac{R_i^{\alpha\beta}R_j^{\alpha\beta}}{R^{\alpha\beta}}\rangle - \frac{NkT\delta_{ij}}{V}$$

Summation over all distinct pairs of atoms $<\alpha\,\beta>$ is performed. $R^{\alpha\beta}$ is the interparticle distance of the pair under consideration, and $R_i^{\alpha\beta}$ denotes the *i*th

Cartesian component of the vector $\vec{R}^{\alpha\beta} \equiv \vec{R}^{\alpha} - \vec{R}^{\beta}$. The symbol < > denotes a thermal average. The term $-NkT\delta_{ij}/V$ is the kinetic contribution to the stress, which originates in the additive term –NkT ln $V$ in the free energy. Here, we calculated the average normal stress $\sigma = (\sigma_{xx} + \sigma_{yy})$ and the shear stress $\sigma_{xy}$.

The variations of normal stress during soliton formation and soliton annihilation are shown in Figure 5a and 5c respectively. The normal stress $\sigma$ decreases as more and more solitons are formed, and vice versa. Also it is clear that the normal stress in the colloidal crystal changes sharply once soliton(s) are formed and annihilated. Interestingly, we find that the extent of the change of the normal stress value is roughly inversely proportional to the change of the number of solitons.

In the structural transformation that leads to soliton formation, we identify the precursory sliding transition before the start of the transformation. The sliding centre appeared and disappeared from time to time until the first soliton was formed. In the plot of the normal stress variation, we also find the sudden stress change which corresponds to the existence of the sliding centre in the colloidal crystal. The stress value goes down to a lower level for a small period of MC time, and then rapidly returns to the original level of the stress.

We also record the shear stress $\sigma_{xy}$ of these two transformations, which is shown in Figure 6a and 6b. It is obvious that the fluctuations of the shear stress in the unstrained crystal, relative to that in the strained crystal have a very small amplitude. For the precursory sliding transition, the sudden large changes of the shear stress are also found before the continuous fluctuations due to the occurrence of the soliton formation.

A more quantitative account of the transition kinetics is given in Figure 5a and b for the transition from the crystal without solitons to the crystal with standing strain wave pattern, and in Figure 5c and d for the reverse transition. These figures illustrate how we can obtain a probability distribution $P(\sigma)$ that a stress $\sigma$ occurs during the transformation. The very high peaks of $P(\sigma)$ at the low end and at the high end of the distribution represent the initial and final state, respectively, while the 6 intermediate peaks (labelled as 1,2, …,6 in Figure 5b) represent the intermediate plateaus, in Figure 5a; these plateaus are metastable intermediate states, and are characterized by much larger fluctuations than both the initial and final state, respectively. The gaps G1, G2, …, G6 in between the peaks of the stress distribution $P(\sigma)$ represent the relatively fast transitions from one metastable state to the next one. Of course, also these “relatively fast” transitions are still rather slow processes, on the time scale (Monte Carlo step per particle) the time needed for these transitions is in between 12500 MCS (Gap1) and 48000 MCS (Gap5). The “lifetime” of the intermediate metastable plateaus in Figure 5a is still distinctly larger, e.g. for plateau 1 it is about 760 000 MCS! Figure 5c and d show similar phenomena for the reverse process. The estimated lifetimes of these intermediate metastable states, as well as the time that the transitions between them take, are collected in Tables 1 and 2 for the soliton formation and the soliton annihilation, respectively. Since all these times are so large, our statement that we deal with very slow and thermally activated dynamics, and hence a Monte Carlo modelling of the stochastic transitions that occur in such a system is adequate, is in fact fully corroborated by our data.

## Conclusion

We have used the Monte Carlo method to simulate the long time non-equilibrium dynamics of a confined model two-dimensional colloidal crystal. A coarse-grained description of the dynamics of these strained lattice systems is given by use of a combination of superimposed configurations and the behavior of the normal and shear stresses. By this method we have shown that the conditions under which the structural transformations occur may be readily identified by these means and that an understanding of these novel dynamical properties should be important to the experimental control of the structure of confined colloidal crystals, self-assembly of nano- and meso- scaled particles, mono-layers on stepped surfaces, etc.

In order to be able to extend our study to related systems described by different dynamics, such as adsorbed atoms on stepped crystal surfaces, Molecular Dynamics (MD) simulations are currently being performed and detailed analysis are being carried out, in order to ascertain the influence of the dynamics on the progress of the layer transitions in strained two-dimensional crystals with confinement. These MD result will be presented elsewhere.

## Acknowledgements

This research was partially supported by the Deutsche Forschungsgemeinschaft Project TR6/C4. We thank H. J. Elmers for a useful discussion.

## Supporting Information Description

The molecular movies of formation (Part1-3) and annihilation (Part1-3) of solitons and standing strain wave superstructures in a two-dimensional colloidal crystal, and the molecular movies of solitons and standing strain wave superstructures without transition (Part1-2).

## Figure Captions

**Figure 1** Internal stress $\sigma = \sigma_{xx} - \sigma_{yy}$ (in LJ units) in the confined crystalline strip plotted vs. $\Delta$, for the case of a system started with $n_x = 30$, $n_y = 108$ (filled symbols) and a system started with $n_x = 29$; $n_y = 108$ (open symbols) plus the appropriate extra particles per row, as described in the text. Open circles represent the stress of the system before the soliton annihilation, and Open triangles represent the stress of the system after the soliton annihilation. With *D* larger than 1.5, the soliton annihilation does not occur. Curves are guides to the eye only.

The upper insert shows a schematic sketch of our geometry: we study a system of size $D$ in the $x$-direction and $L_y$ in the $y-$direction, apply a periodic boundary condition along the $y-$axis, while the boundary in the $x-$direction is created by two rows of fixed particles (shaded) on the ideal positions of a perfect triangular lattice with lattice spacing $a$ at each side. In the fully commensurate case, $D = n_x a\sqrt{3}/2$. The open circles represent the first row of mobile particles adjacent to each wall.

**Figure 2** The origin of soliton lattice in two-dimensional colloidal crystal: (a) In perfect crystal, all the particles are located at the bottom of the potential energy wells and the system is fully commensurate. (b) The distance between structured walls is reduced. The number of particle row is reduced by one, and the particles from the missing row compete for the space with the particles in the remaining rows. Some of the particles should move out of the bottom of potential energy wells, and one of them (black) is even located at the top of the potential energy barrier. The system becomes incommensurate. (c) The zoom-in superimposed configurations show two solitons

near to the structured wall. The back square dots represent static wall particles.

**Figure 3** Superimposed configurations with a misfit of Δ = 2.0. The numbers indicate Monte Carlo (MC) steps after the start of the simulation run. Each snapshot consists of 100 individual configurations, and each configuration is generated every 100 MC moves. The top and bottom two rows of static particles are the structured walls which provide the confinement. The arrows indicate the direction of the motion of the sliding centre which leads to the formation of a soliton. The ellipses indicate the newly formed soliton after the sliding transition in the two-dimensional colloidal system. The movie (see the supplementary information) is generated from consecutive superimposed configurations from this Monte Carlo run.

**Figure 4** Superimposed configurations with a misfit of Δ = 1.5. The Monte Carlo (MC) simulation started with a system started with $n_x = 29$; $n_y = 108$ plus the appropriate extra particles per row. The numbers indicate the number of MC steps after the start of the simulation run. After a very short MC run, the extra particles lead to the formation of solitons and a strain wave structure. The solitons are however not stable because the misfit Δ is too small to stabilize the strained system. As a result the annihilation of the solitons occurs. The arrows and ellipses indicate the sliding direction and location of annihilated solitons. (The molecular movie can be found as supplementary information)

**Figure 5** (a) Time evolution of the stress $\sigma$ during the transition from the crystal without solitons to the crystal exhibiting soliton stair cases causing the standing strain wave pattern, showing the assignment of 6 plateaus 1,2,3,4,5,6 and gaps G6 (between plateaus 6 and 5), G5 (between plateau 5 and 4, etc.) (b) shows the resulting stress distribution, $P(\sigma)$, where plateaus show up as peaks, and the minima in between (gaps G1, G2, $\cdots$) represent the transitions between these metastable plateaus. Note that the area of the peaks in $P(\sigma)$ can be taken as a measure of the lifetime of the state which the peak belongs to. (c) and (d) show the case of reverse transition.

**Figure 6** (a) The variation of shear stress during the soliton formation and (b) the variation of shear stress during the soliton annihilation.

## Table Captions

**Table I:** Label of peak (upper part) or gap (lower part) for the forward transition, shown in left column; associated area of $P(\sigma)$, middle column; estimated lifetimes of the metastable states (right column, upper part) or passage times to move from one plateau to the next (right column, lower part)

**Table II**: Same as Table I, but for the reverse transition.

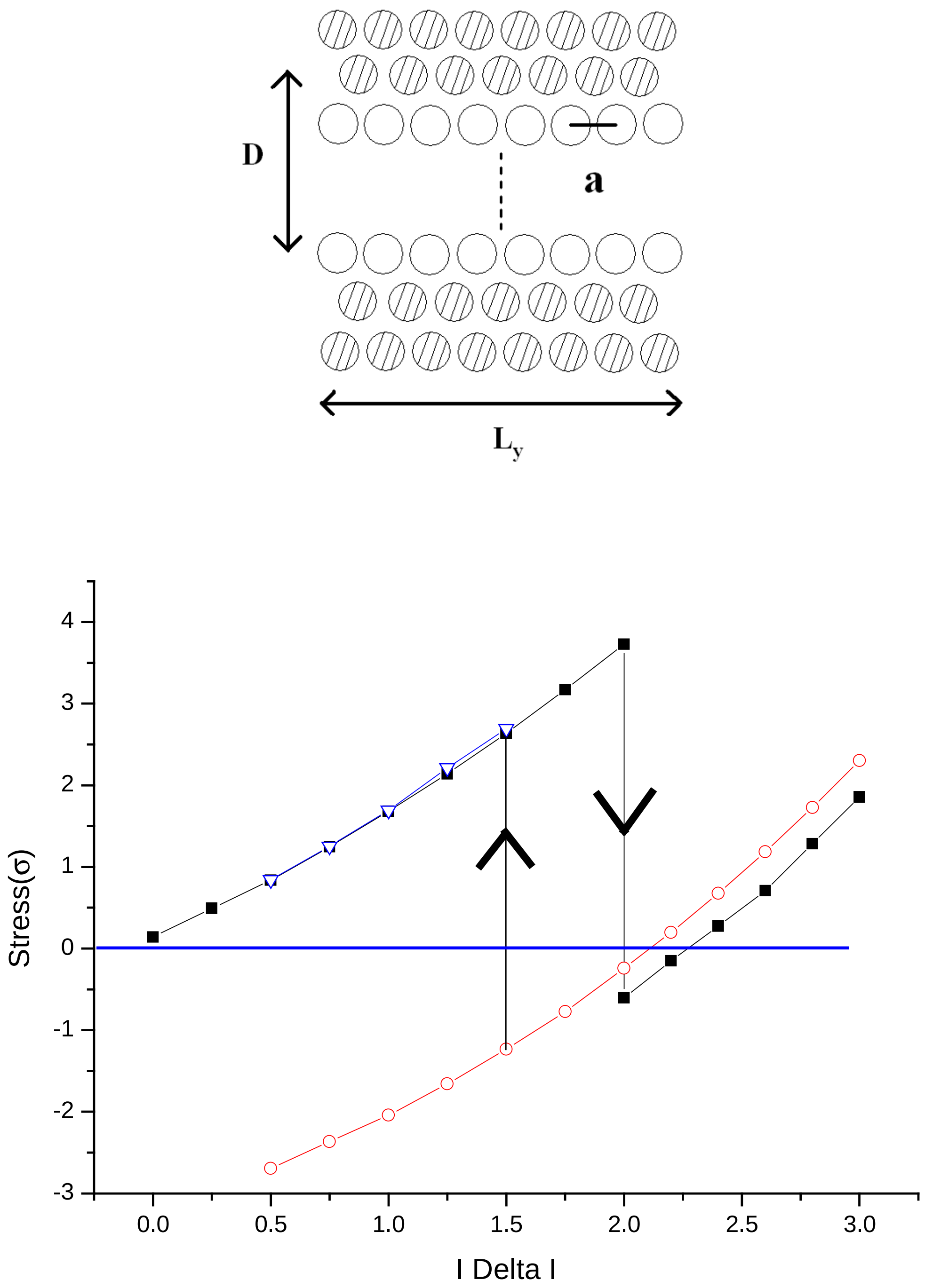

Figure 1

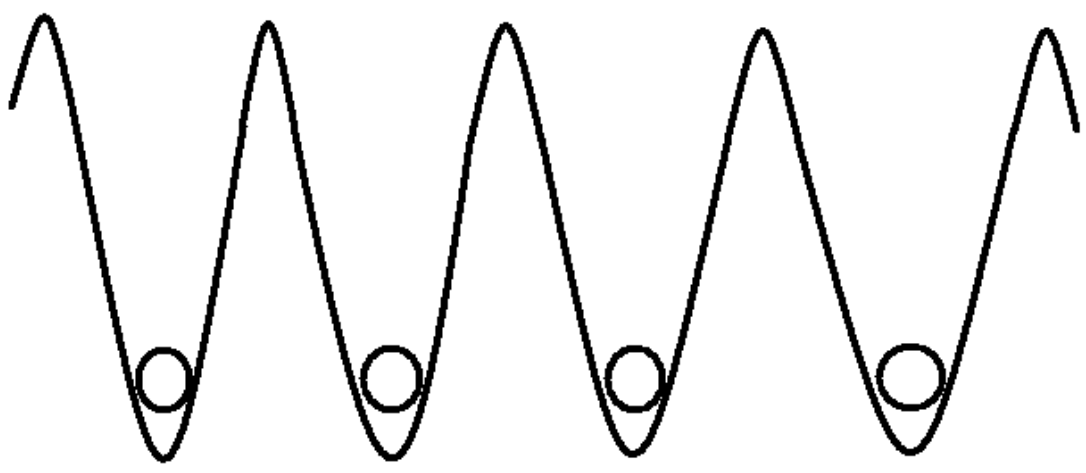

(a)

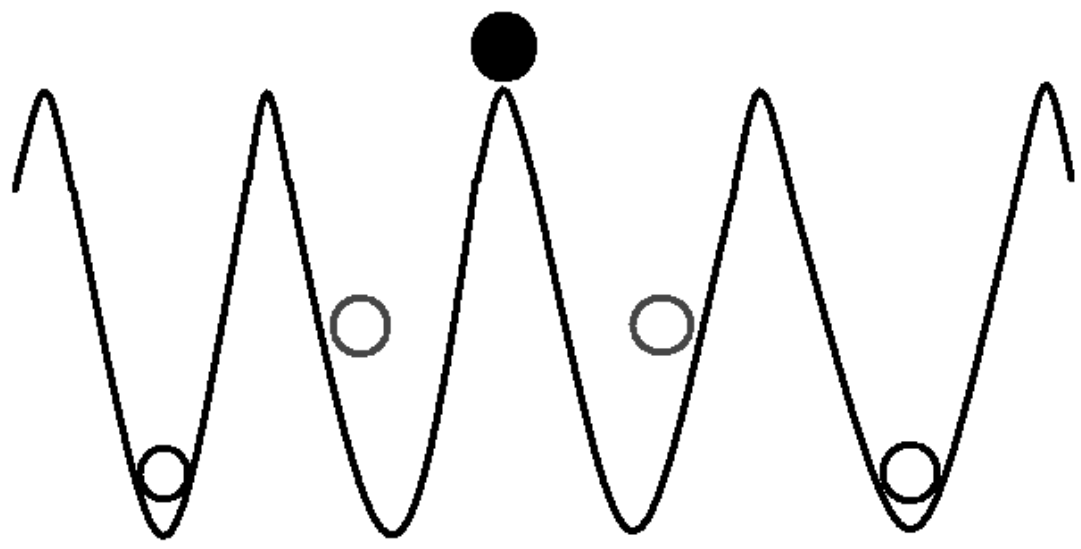

(b)

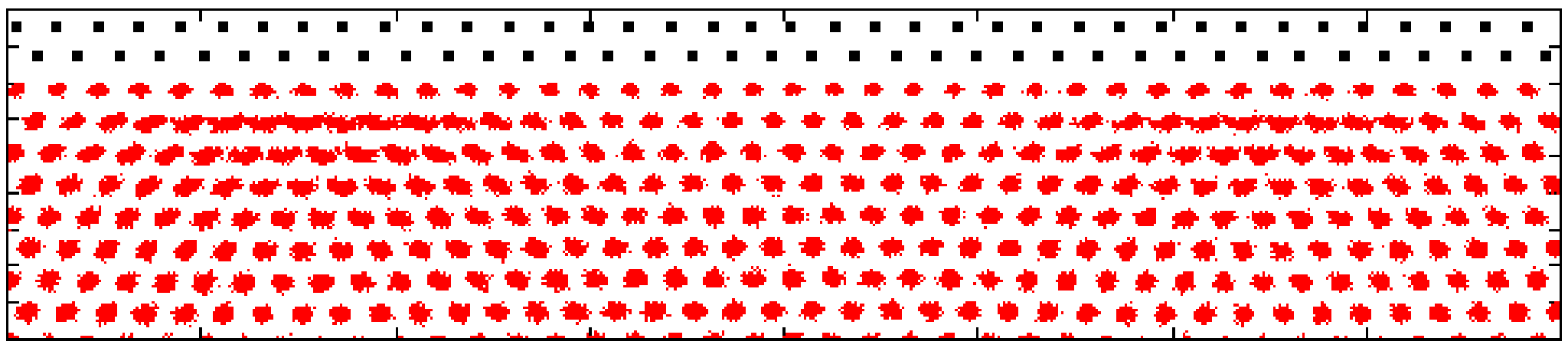

(c)

Figure 2

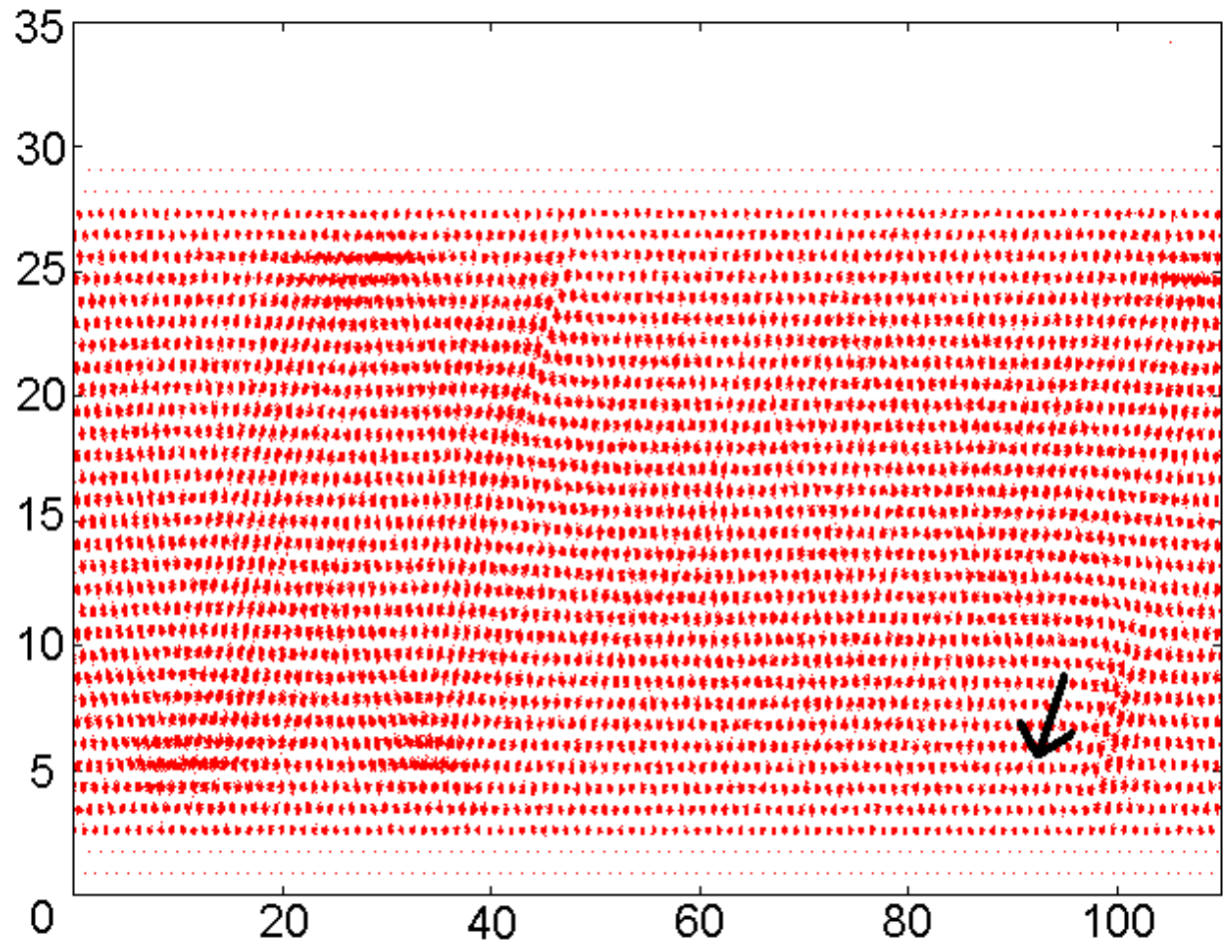

(a) 7700 MC steps

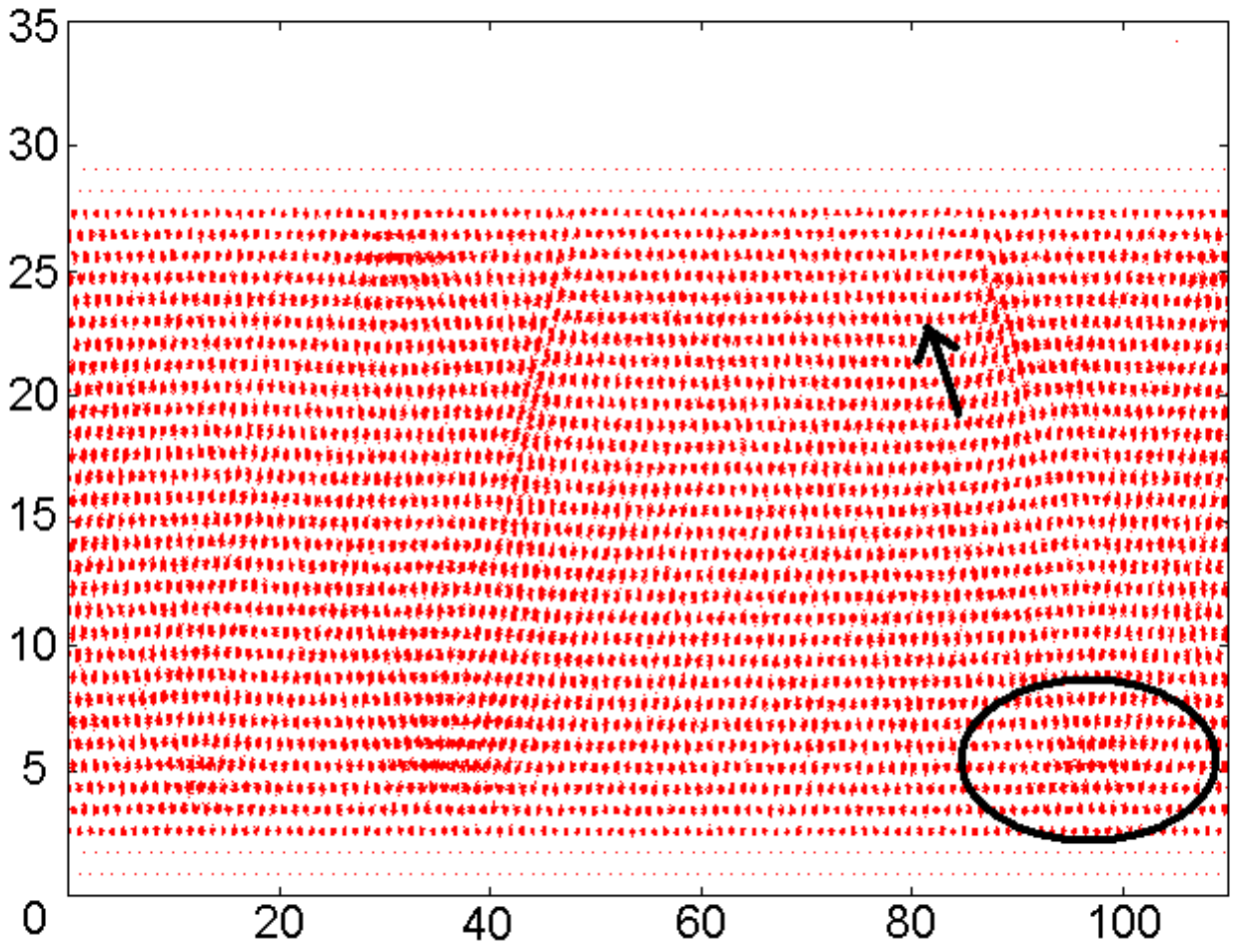

(b) 25000 MC steps

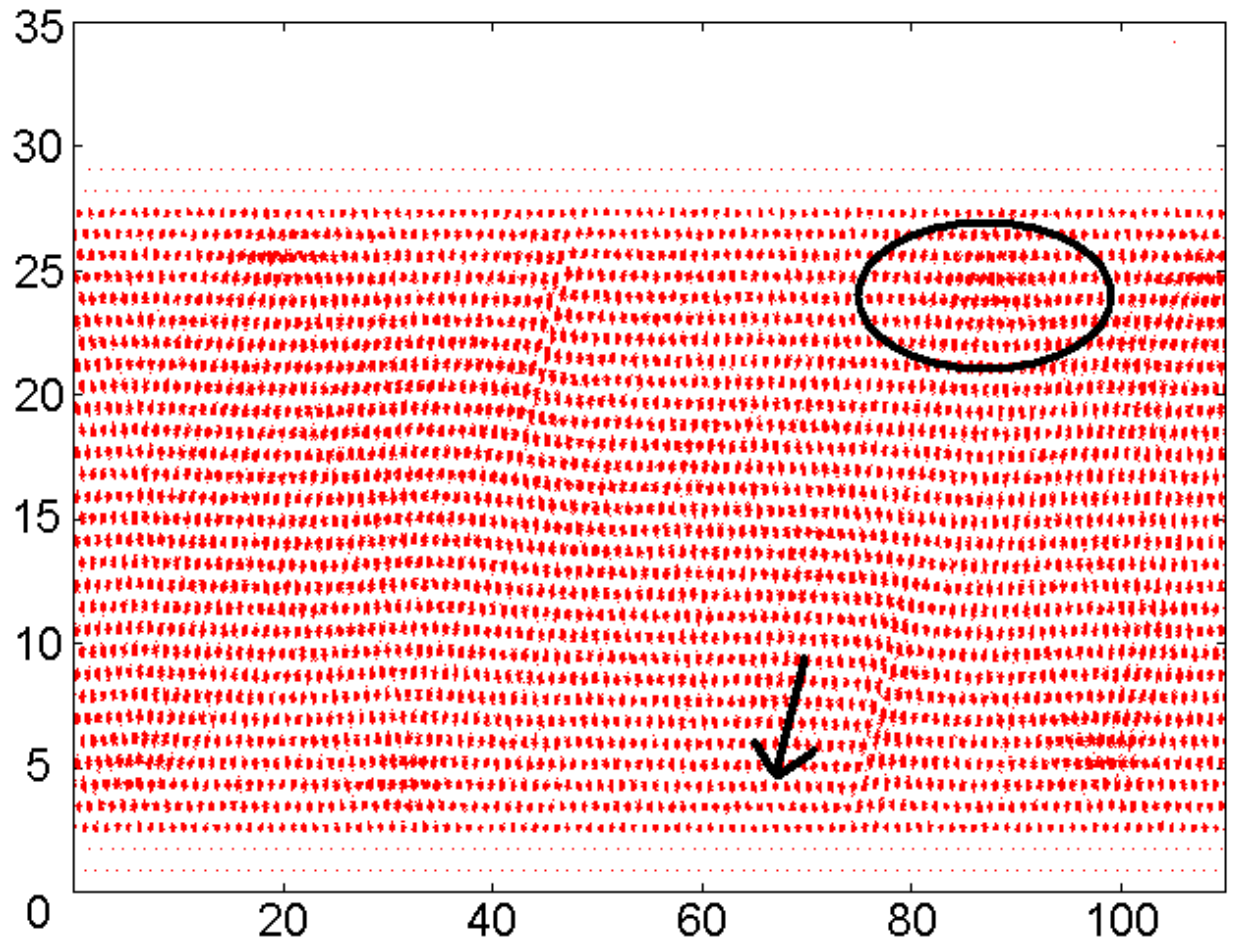

(c) 40500 MC steps

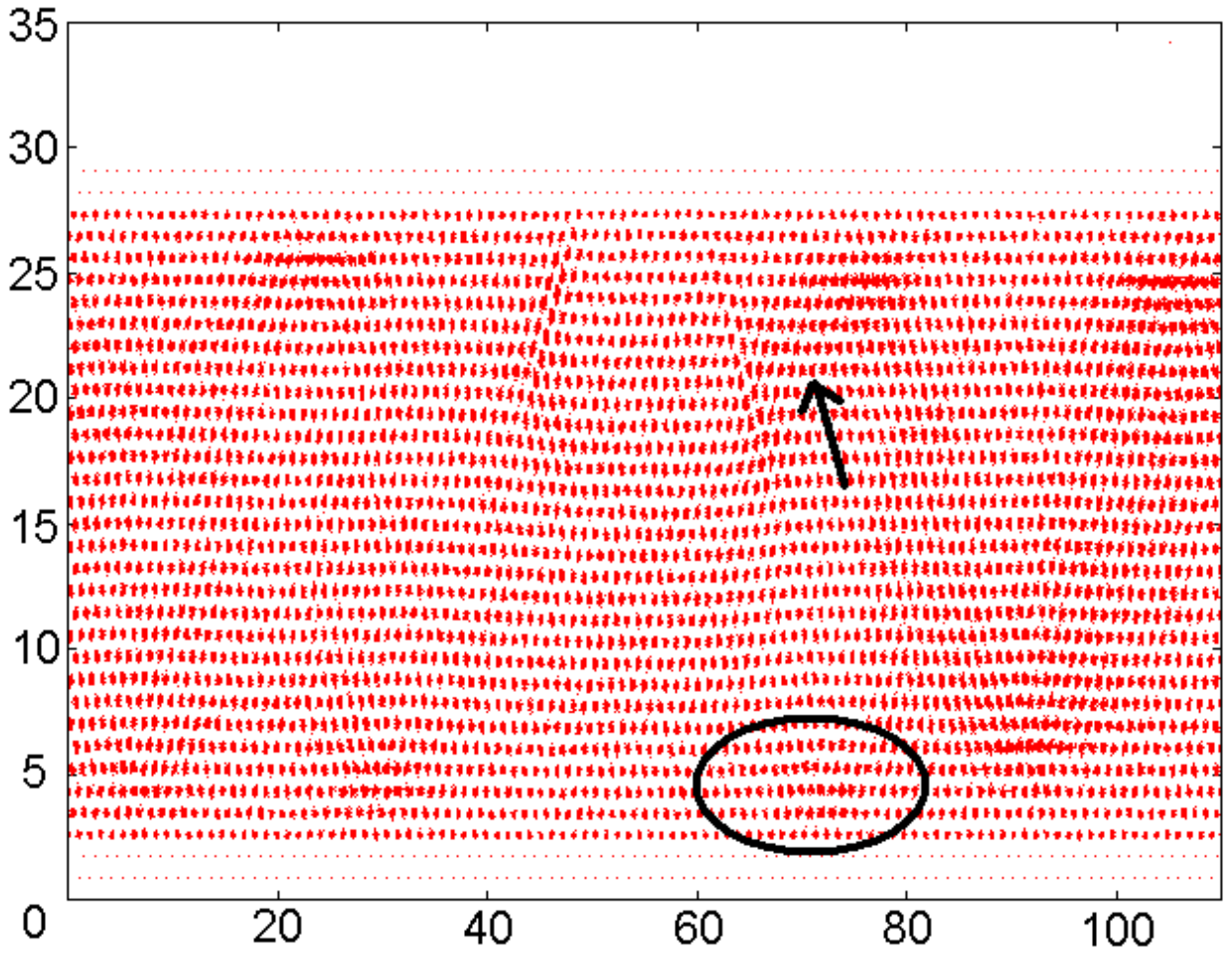

(d) 65500 MC steps

Figure 3

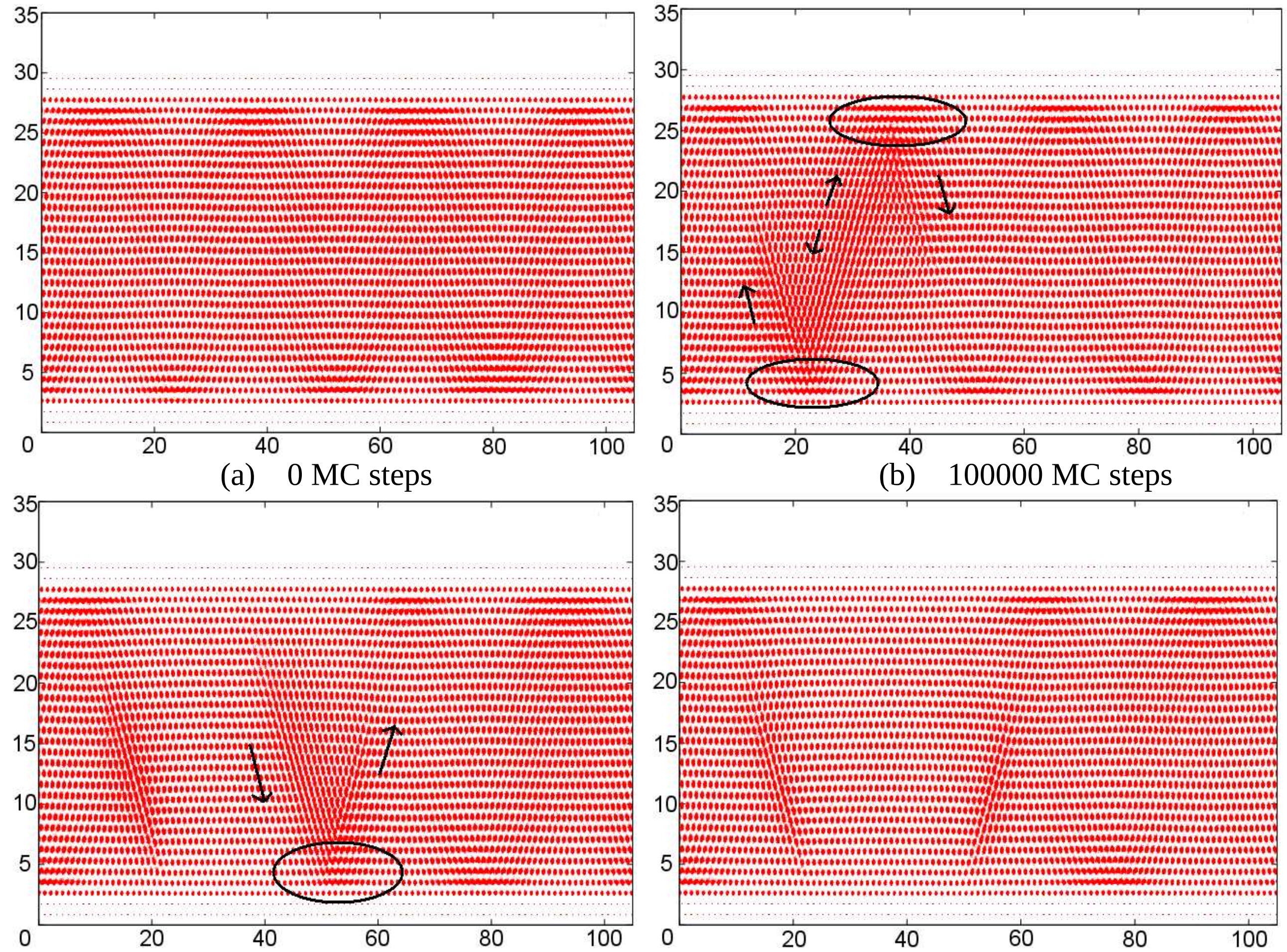

(a) 0 MC steps

(b) 100000 MC steps

(c) 200000 MC steps

(d) 300000 MC steps

Figure 4

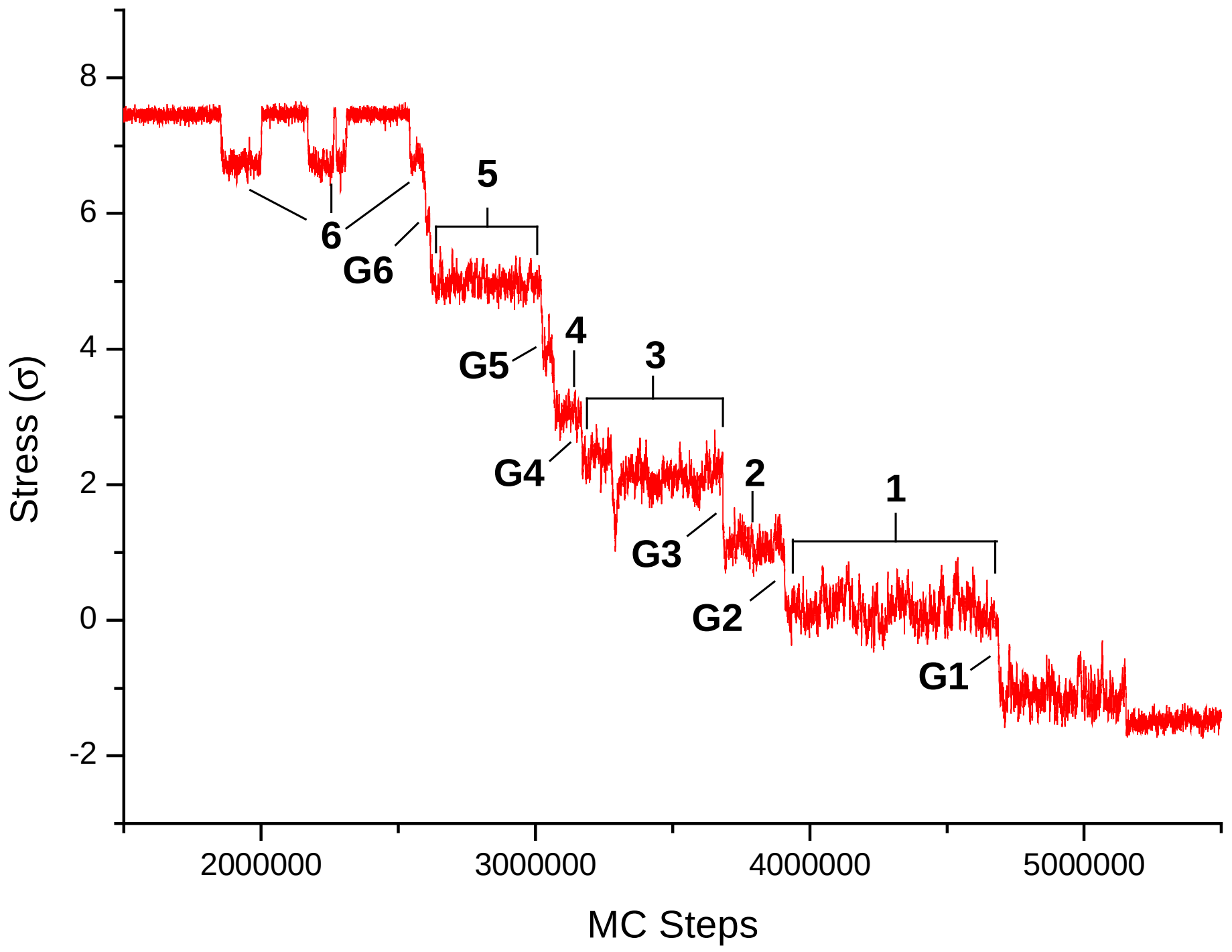

(a)

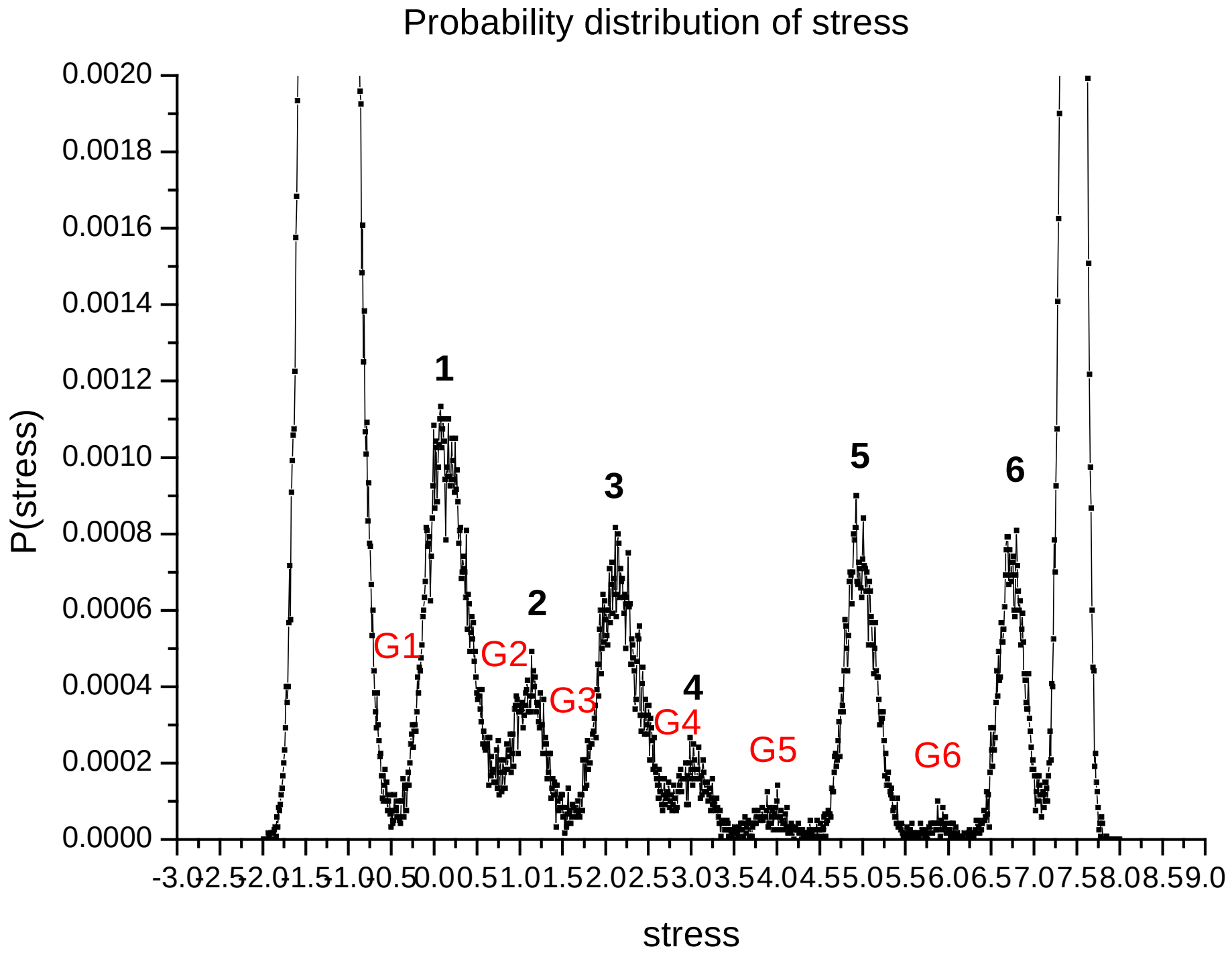
Probability distribution of stress
P(stress)
stress
0.0020
0.0018
0.0016
0.0014
0.0012
0.0010
0.0008
0.0006
0.0004
0.0002
0.0000
-3.02.52.01.51.00.50.00.51.01.52.02.53.03.54.04.55.05.56.06.57.07.58.08.59.0
1
2
3
4
5
6
G1
G2
G3
G4
G5
G6

(b)

Figure 5

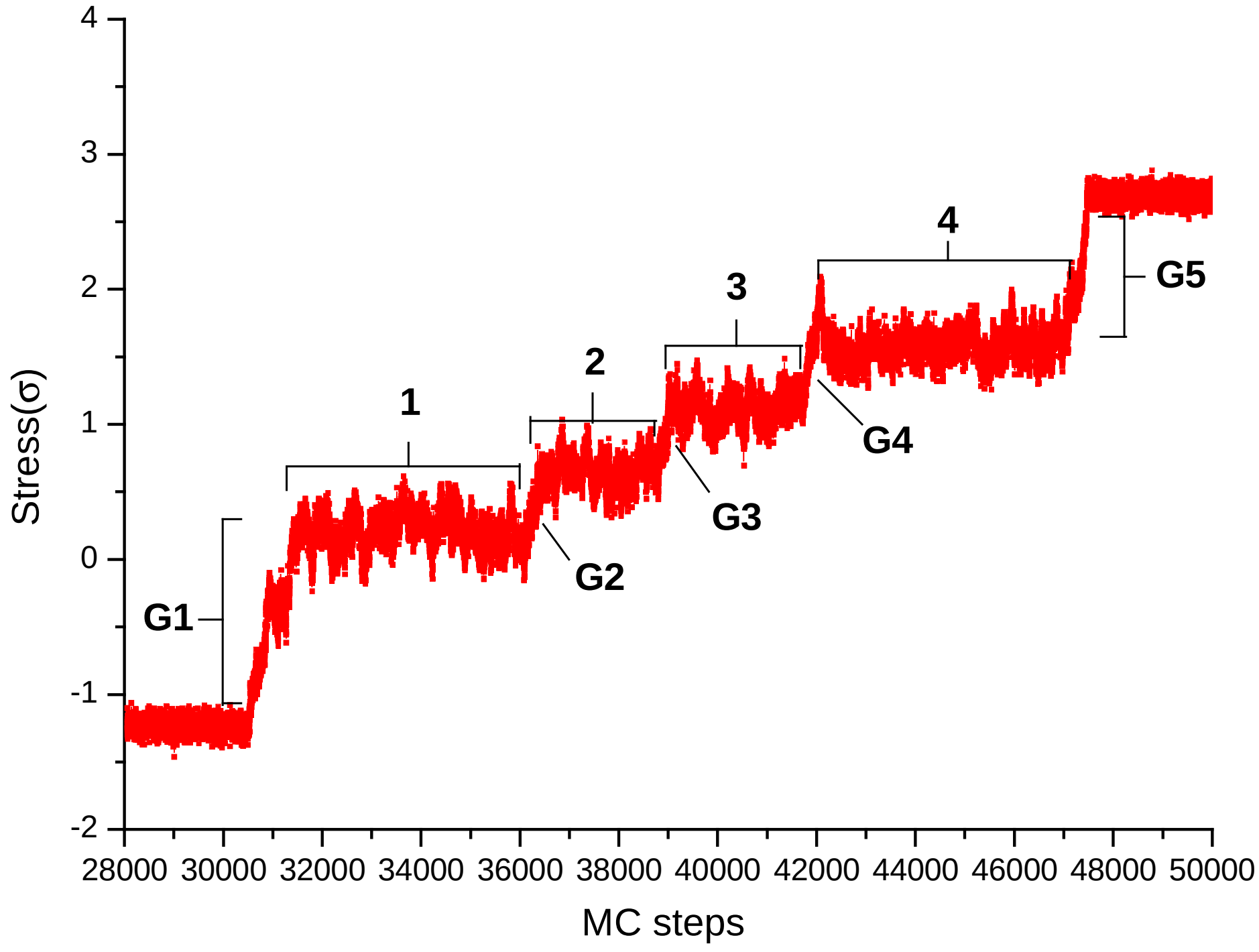

Stress(σ)
MC steps
4
3
2
1
0
-1
-2
28000 30000 32000 34000 36000 38000 40000 42000 44000 46000 48000 50000
G1
1
2
3
4
G2
G3
G4
G5

(c)

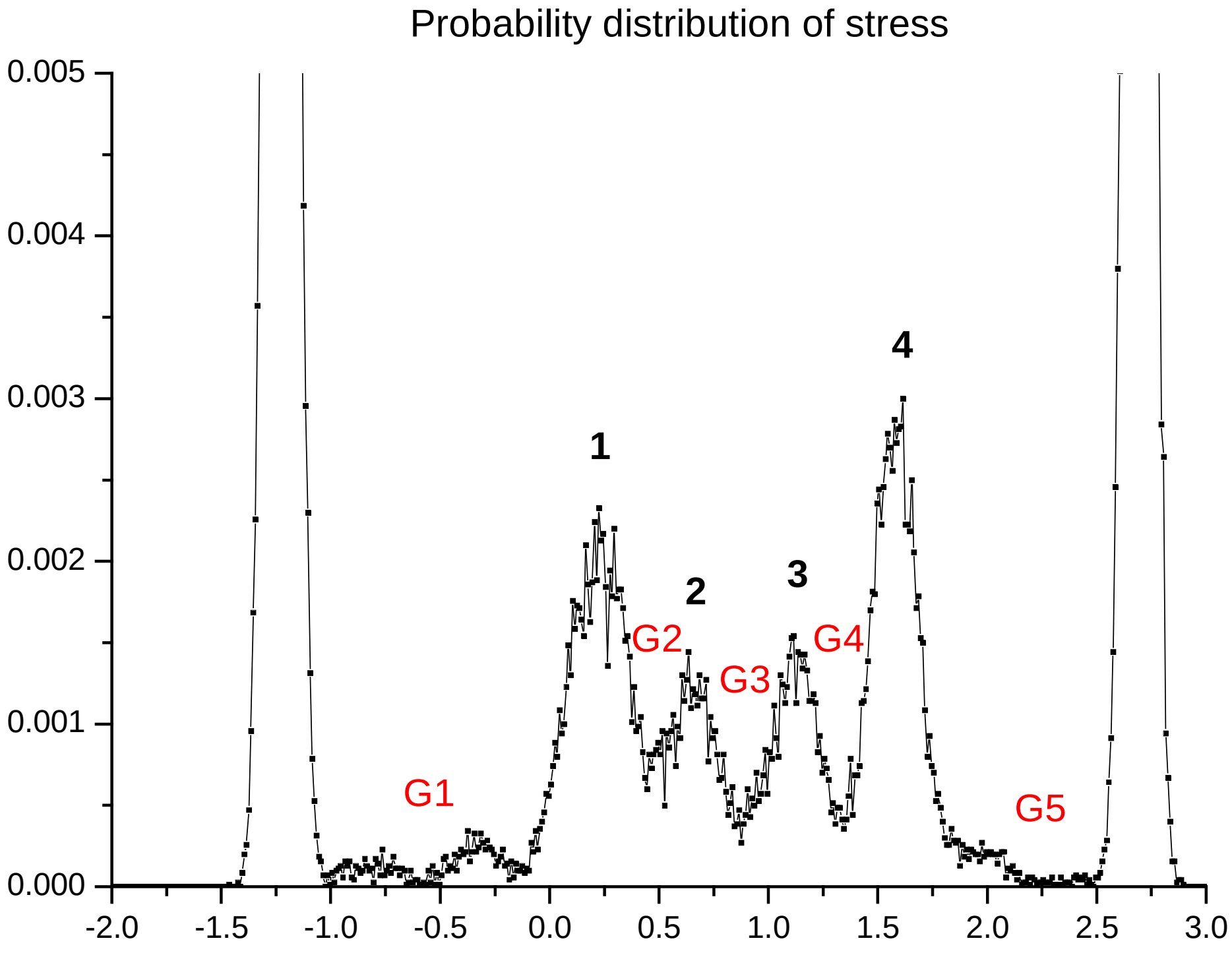

Probability distribution of stress
0.005
0.004
0.003
0.002
0.001
0.000
-2.0 -1.5 -1.0 -0.5 0.0 0.5 1.0 1.5 2.0 2.5 3.0
1
2
3
4
G1
G2
G3
G4
G5

(d)

Figure 5

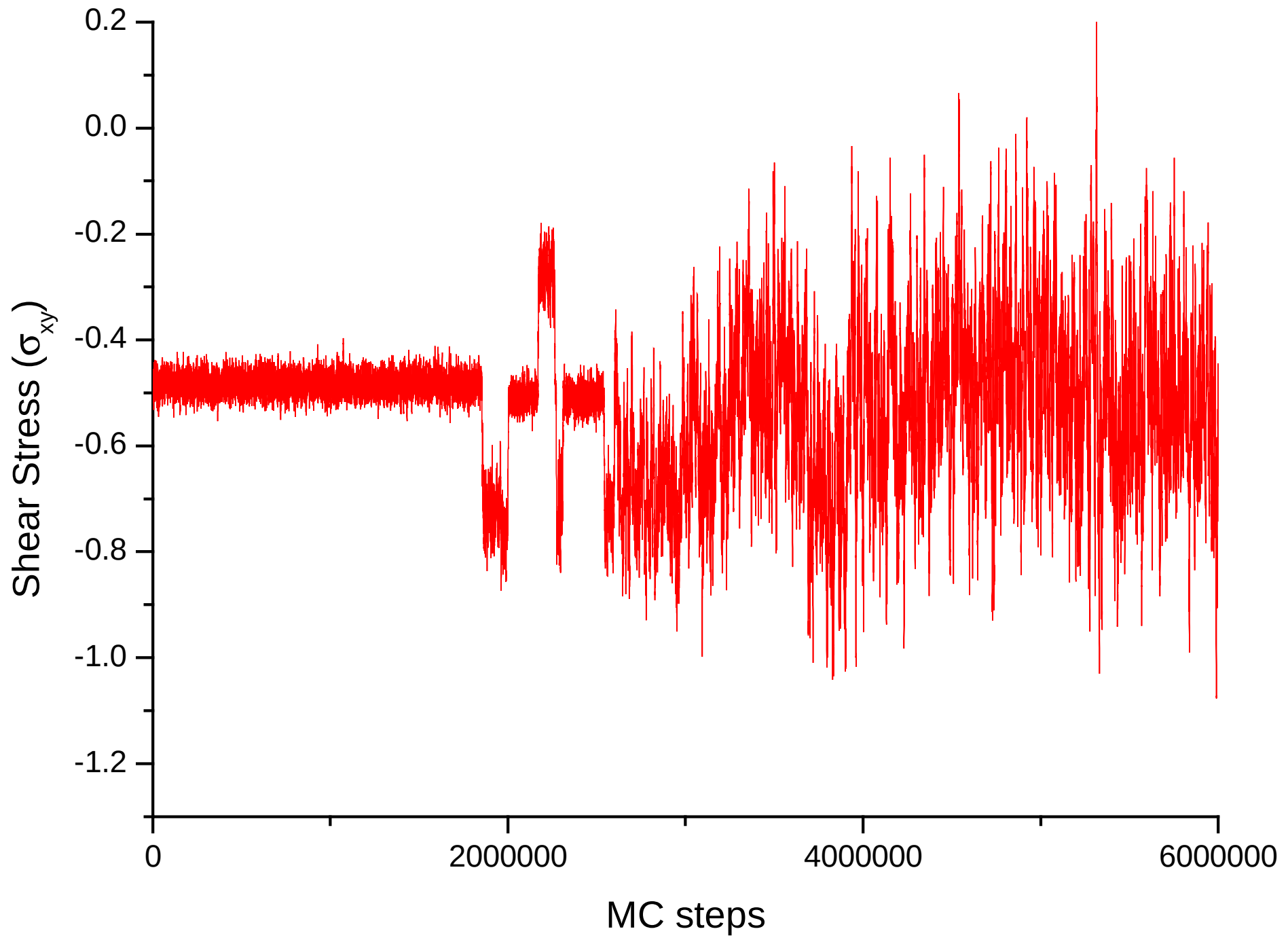

(a)

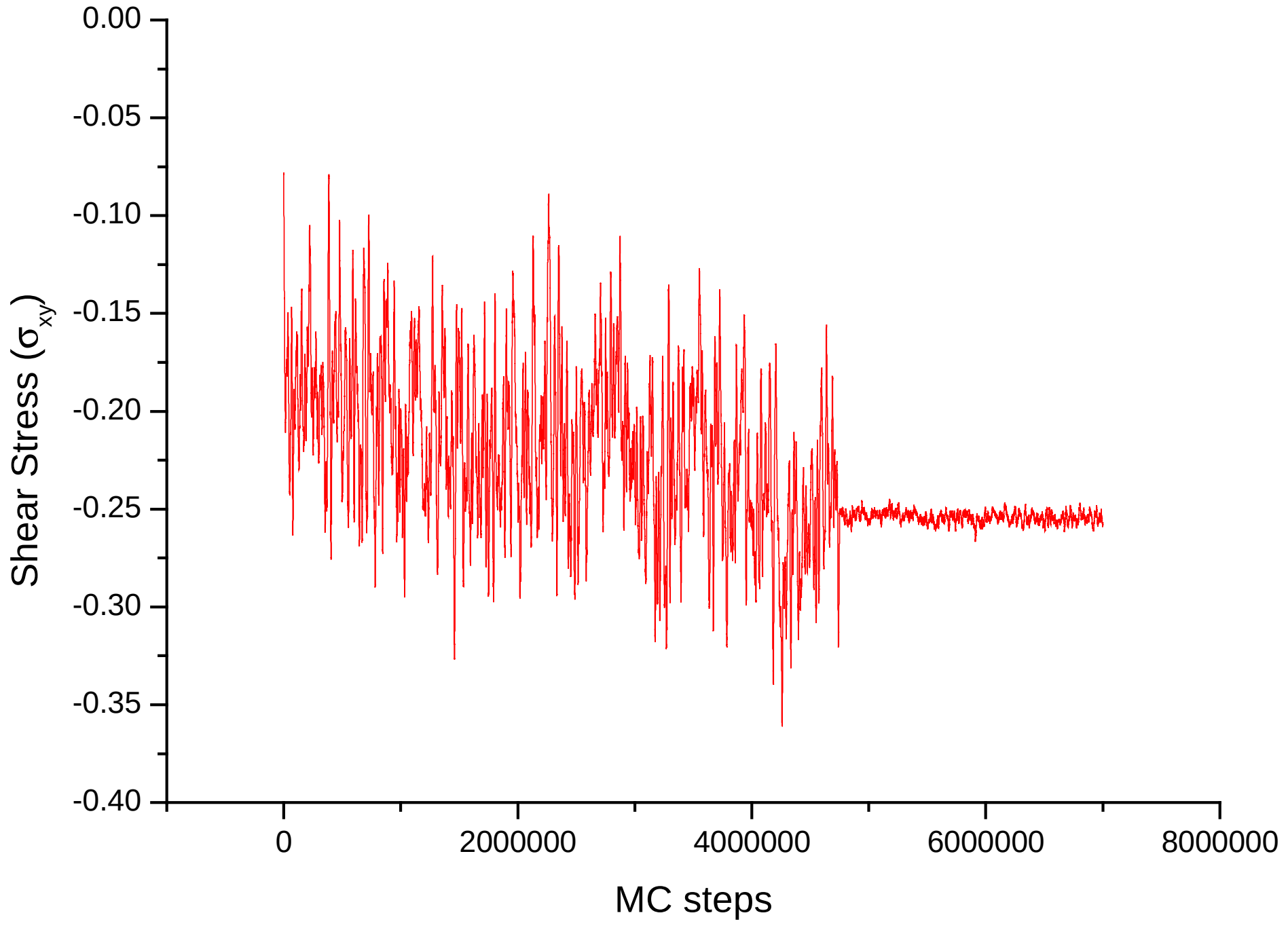

(b)

Figure 6

| Peak | Area | Estimated MC steps (Number obtained from Stress vs MCS) |
|---|---|---|
| 1 | 0.06349 | 761500 (~763800) |
| 2 | 0.01712 | 205300 (~202100) |
| 3 | 0.0393 | 471300 (~478400) |
| 4 | 0.00763 | 91400 (~87400) |
| 5 | 0.03361 | 403200 (~402600) |
| 6* | 0.02607 | 312700 (~316100) |

| Gap | Area | Estimated MC steps |
|---|---|---|
| 1 | 0.00104 | 12500 |
| 2 | 0.00154 | 18500 |
| 3 | 0.00235 | 28300 |
| 4 | 0.00222 | 26600 |
| 5 | 0.004 | 48000 |
| 6 | 0.00201 | 24100 |

Table 1

| Peak | Area | Estimated MC steps (Number obtained from Stress vs MCS) |
|---|---|---|
| 1 | 0.06617 | 463200 (~479400) |
| 2 | 0.03329 | 233100 (~234200) |
| 3 | 0.03753 | 262700 (~280100) |
| 4 | 0.06981 | 488700 (~514100) |

| Gap | Area | Estimated MC steps |
|---|---|---|
| 1 | 0.01154 | 80800 |
| 2 | 0.00333 | 23300 |
| 3 | 0.00397 | 27800 |
| 4 | 0.00448 | 31300 |
| 5 | 0.00771 | 54000 |

Table 2